\documentclass[12pt]{article}
\usepackage{amsfonts}
\usepackage{graphicx}
\usepackage{amsmath}
\usepackage{float}
\usepackage{amssymb}
\providecommand{\U}[1]{\protect\rule{.1in}{.1in}}
\textwidth=6.5in \hoffset=-.75in \voffset=-1in \numberwithin{equation}{section}
\numberwithin{figure}{section}

\begin{document}

\begin{titlepage}
\bigskip \begin{flushright}
\end{flushright}
\vspace{1cm}
\begin{center}
{\Large \bf {Supergravity Solutions of Two M2 Branes}}\\
\vskip 1cm
\end{center}
\vspace{1cm}
\begin{center}
A. M.
Ghezelbash{ \footnote{ E-Mail: masoud.ghezelbash@usask.ca}}, R. Oraji
{ \footnote{ E-Mail: rao519@mail.usask.ca}}
\\
Department of Physics and Engineering Physics, \\ University of Saskatchewan,
Saskatoon, Saskatchewan S7N 5E2, Canada\\
\vspace{1cm}
\end{center}
\begin{abstract}

We present an exact analytic class of solutions for a system of two membranes in eleven-dimensional supergravity. One brane in the system is completely localized along the overall and relative transverse coordinates while the other brane in the system is localized only along the overall transverse coordinates. The membrane configuration preserves four supersymmetries. Moreover we find some approximate solutions for the system of two membranes with a Bianchi space as the overall transverse space to both membranes. All supergravity solutions preserve 1/8 of the supersymmetry. Upon dimensional reduction, the solutions provide intersecting configurations of three D-branes in type IIA supergravity. 

\end{abstract}
\bigskip
\end{titlepage}\onecolumn

\bigskip

\section{Introduction}
\label{sec:Introduction}

All different brane solutions in eleven-dimensional supergravity are the classical solitons of M-theory, so there has been lot of interest and achievement in constructing and classifying the brane solutions of eleven-dimensional supergravity \cite{Tsey}-\cite{gauntlettetal3}. These solutions generate the p branes and the solitons of string theory that in turn make a way to understand the quantum theory of black holes. The black hole structure can be considered as some wrapped specific p branes around some compact manifoldes. Moreover the supergravity brane solutions are the systems that provide evidence for the bulk-boundary holography duality \cite{gr1}-\cite{DecouplingLim1}.

One other important feature of the solutions is whether they are stable supersymmetric solutions or not. It turns out that in general, orthogonal intersecting system of M2 and M5 branes preserve some number of supersymmetry and are stable systems \cite{Tsey}. It is also a well known fact that for supersymmetric solutions of supergravity, the metric functions can be obtained in terms of harmonic functions  \cite{Tsey}-\cite{Gaun}. The harmonic functions are on the transverse space to brane systems and related to the saturated BPS bound of the brane systems. 

As long as the near core region of supersymmetic BPS solutions of brane sysytems is of importance in bulk-boundary duality, numerous solutions of orthogonal intersecting M branes have been constructed  
\cite{Tsey}-\cite{gauntlettetal3}, \cite{IT}-\cite{smith}. 

However, to find the brane solutions that are not restricted to the near core region of D4, D5 or D6 branes, one can lift the corresponding brane to an embedded self-dual (or anti-self-dual) space in eleven-dimensional supergravity \cite{hashi}. Upon dimensional reduction, the M-brane solutions with self-dual (or 
anti-self-dual) transverse space yield systems of p branes that are not restricted to the near core of D4, D5 or D6 branes. Instead the solutions are on near core of lower dimensional branes, such as D2 where its world-volume theory enjoys remarkable properties such as renormalizability. Moreover, the lower dimensional brane is always fully localized in the world volume of higher dimensional brane and the system preserves some suspersymmetries. 
The method works very well for a single M2 or M5 brane, as by assuming an  ansatz for the eleven dimensional metric, the supergravity equations of motion  reduce to one partial differential equation that can be solved mainly by numerical methods after imposing the proper boundary conditions \cite{CGMM2}-\cite{Rahim}. 

In this paper, we consider a system of two intersecting M2 branes and find a class of  
exact analytic solutions to supergravity equations of motion.
The equations of motion consist of two coupled partial differential equations. The equation of motion for the metric function of one membrane depends on the metric function of second membrane in the system.  
Upon dimensional reduction, we get a system of three D branes in type IIA supegravity. We explicitly calculate the number of preserved supersymmetries by finding the solutions to the Killing spinor equation and show the system of two M2 branes preserves four supersymmetries.

The outline of the article is as follows. In section \ref{sec:SupergravitySolutions}, we
briefly discuss
the eleven dimensional supergravity equations of motion and 
the ansatz for the eleven-dimensional metric.
In section \ref{sec:TN}, we present our exact analytical solutions to the two coupled differential equations for the metric functions of two M2 branes. We find a second set of solutions to the equations of motion by analytically continuing the separation constant that appear in the first exact analytical solutions.
In sections \ref{sec:Bianchi}, we consider the four dimensional triaxial Bianchi IX space as a part of the overall transverse space to two membranes.  Although it is very unlikely to find an analytical solutions to equations of motion, however we find approximate analytical solutions for the membrane metric functions.
In section \ref{sec:Supersymmetry}, we find the explicit solutions to the Killing spinor equation for the system of two M2 branes. 
We show all of the solutions presented in sections \ref{sec:TN} and \ref{sec:Bianchi} preserve four supersymmetries. In section \ref{sec:Summary}, we wrap up the article by concluding
remarks and future possible research directions.

\section{Supergravity solutions}
\label{sec:SupergravitySolutions}
The equations of motion for the bosonic fields of $D=11$ supergravity 
are given by \cite{DuffKK}
\begin{align}
R_{MN}-\frac{1}{2}g_{MN}R  &  =\frac{1}{3}\left[  F_{MPQR}F_{N}
^{\phantom{N}PQR}-\frac{1}{8}g_{MN}F_{PQRS}F^{PQRS}\right], \label{GminGG}\\
\nabla_{M}F^{MNPQ}  &  =-\frac{1}{576}\varepsilon^{M_{1}\ldots M_{8}
NPQ}F_{M_{1}\ldots M_{4}}F_{M_{5}\ldots M_{8}}, 
\label{dF}
\end{align}
where 
$F_{M_1M_2M_3M_4}=4\partial_{[M_1} A_{M_2M_3M_4]}$ is the field strength of the three form gauge field $A_{M_1M_2M_3}$. The solutions to equations of motion (\ref{GminGG}) and (\ref{dF}) may preserve partial supersymmetry if there exists non-trivial Killing spinors that satisy the Killing spinor equation. The Killing spinor equation indicates that the supersymmetric variation of the gavitino field vanishes.

We consider a system of two M2 branes in which the branes are located in $\rho=0$ and $r=0$ with the metric \cite{Youam}
\begin{equation}
ds_{11}^{2}=H_1^{\frac{1}{3}}H_2^{\frac{1}{3}}\left( -\frac{dt^{2}}{H_1H_2}+\frac{1}{H_2}(dx_{1}^{2}+dx_{2}^{2}) +\frac{1}{H_1}(dy_{1}^{2}+dy_{2}^{2})+ d\mathfrak{s}_{4}^{2}(r)+{d{\rho}}^2+{\rho}^2 {d\eta}^2 \right),  \label{ds11genM2}
\end{equation}
where $0<\rho<+\infty$ and $0\leq\eta<2\pi$. Moreover, we consider the dependence of the metric functions on transverse coordinates as  
$H_1=H_1(x_1,x_2,\rho,r)$
and $H_2=H_2(\rho,r)$, so the metric functions $H_1$ and $H_2$ depend on the overall transverse coordinates $\rho, r$ while $H_1$ depends also on the relative transverse coordinates $x_1, x_2$. This means the first brane is completely localized on transverse directions while the second brane is delocalized along the relative transverse directions. We notice that the delocalization of a brane in a brane configuration is necessary for the decoupling limit of the theory living on the other brane worldvolume \cite{Youam}. 
We also show (in section \ref{sec:Supersymmetry}) the presence of self-dual (or anti-self dual) space $d\mathfrak{s}_{4}^{2}(r)$ in transverse space to both M2 branes implies partial preserved supersymmetry for the system. In this regard, in section \ref{sec:TN}, we consider the four dimensional self-dual transverse space $d\mathfrak{s}_{4}^{2}(r)$ in (\ref{ds11genM2}) as
\begin{equation}
d\mathfrak{s}_{4}^{2}(r)=V(r)({dr}^2+r^2({d\theta}^2+\sin^2\theta {d\phi}^2) )+\frac{{(d\psi+\omega(\theta)d\phi)}^2}{V(r)},
\label{H1H2H}
\end{equation}
where $
V(r)=1+\frac{n}{r} $ and $\omega(\theta)=n \cos \theta$.
The range of coordinates are, $0<r<+\infty$, $0\leq\psi<4\pi n$ and $0\leq\phi<2\pi$ where $n >0$.
In section \ref{sec:Bianchi}, we consider the system of two M2 branes (\ref{ds11genM2}) in background of self-dual triaxial Bianchi IX space 
\begin{equation}
d\mathfrak{s}_{4}^{2}(r)=\frac{{dr}^2}{\sqrt{f(r)}}+\frac{r^2}{4}\sqrt{f(r)}\left( \frac{\sigma_1^2}{1-\frac{a_1^4}{r^4}} +\frac{\sigma_2^2}{1-\frac{a_2^4}{r^4}}+\frac{\sigma_3^2}{1-\frac{a_3^4}{r^4}} \right),
\label{BIX}
\end{equation}
where the $SO(3)$ invariant one forms $\sigma_i$'s are 
\begin{subequations}\label{sigma.B.0}
\begin{align}
\sigma_1&=d\psi+\cos \theta d\phi, \label{sigma.B.1} \\
\sigma_2&=-\sin \psi d\theta + \cos \psi \sin\theta d\phi,  \label{sigma.B.2}  \\
\sigma_3&=-\cos \psi d\theta + \sin \psi \sin\theta d\phi, \label{sigma.B.3} 
\end{align}
\end{subequations}
and 
\begin{equation}
f(r)=\prod_{i=1}^3 f_i(r)=\prod_{i=1}^3 \left( {1-\frac{a_i^4}{r^4}} \right)   \cdot
\label{FBianchi}
\end{equation}
In (\ref{FBianchi}), $a_i,\,i=1,2,3$ are three arbitrary constants that can be chosen as $a_1=0$, $a_2=2kb$ and $a_3=2b$. The constant $b$ is positive and $0\leq k\leq 1$. We notice $a_1\leq a_2 \leq a_3$ and coordinate $r$ always should be greater than or equal to $2b$.    
Moreover, we consider
the four-form field strength $F$ in terms of metric functions $H_1$ and $H_2$ as
\begin{align}
F=\frac{1}{2} \left( d(\frac{1}{H_2})\wedge dt \wedge dx_1 \wedge dx_2 + d(\frac{1}{H_1})\wedge dt \wedge dy_1 \wedge dy_2 \right)\cdot
\label{4Field}
\end{align}

Dimensional reduction of the metric \eqref{ds11genM2} (with components $G_{mn}$) together with the four-form field strength (\ref{4Field}) to $D=10$ along one of  the compact spatial coordinates in the transverse space (e.g. $\psi$) yields type IIA supergravity solutions. The fields of type IIA supergravity could be read from the relations 
$G_{\alpha\beta}=e^{-2\Phi /3}\left( g_{\alpha \beta }+e^{2\Phi }C_{\alpha }C_{\beta }\right)$, $G_{\psi\psi}=\nu ^{2}e^{4\Phi /3}$ and $G_{\alpha\psi}=\nu e^{4\Phi /3}C_{\alpha }$, where $\alpha$ and $\beta$ take values in ten dimensions. The winding number $\nu$, shows how many times the membrane wraps around the compact dimension \cite{Townsend1}. In the following sections, we set $\nu=1$. The type IIA RR four-form ${\cal F}_4$ and NSNS three-form field strength $H_3$ are related to (\ref{4Field}) by $F={\cal F}_4+R_0 H_3\wedge d\psi$ where $R_0$ is the radius of circle parameterized by $\psi$. 
\section{Supergravity solutions for a system of two M2 branes}
\label{sec:TN}
\begin{figure}[H]
\centering
 \includegraphics[width=0.5\textwidth]{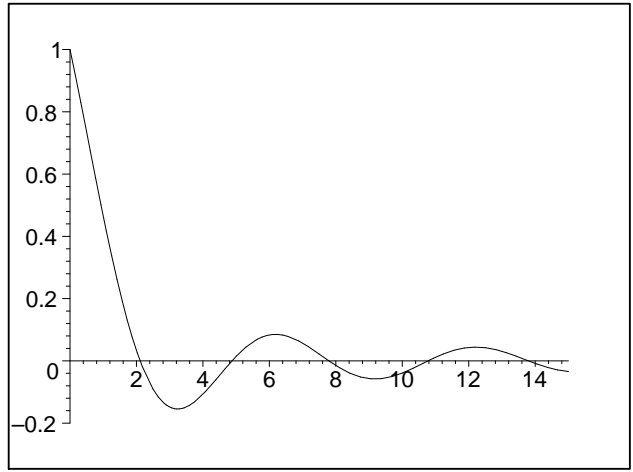}
 \caption{The solution $\mathcal{F}_1(r)$ vanishes far away from the branes. The constants $c$ and $n$ are set to $1$.}
 \label{fig:figure1}
\end{figure}

The metric \eqref{ds11genM2} and the four-form field strength (\ref{4Field}) satisfy the equations of motion (\ref{GminGG}) and (\ref{dF}) contingent on  functions $H_1$ and $H_2$ satisfy the following set of coupled differential equations
\begin{equation}
\frac{1}{\rho}\frac{\partial H_2}{\partial {\rho}} +  \frac{\partial^2 H_2}{\partial {{\rho}^2}} +\frac{1}{V(r)} \left( \frac{2}{r}\frac{\partial H_2}{\partial {r}} +  \frac{\partial^2 H_2}{\partial {{r}^2}} \right)=0,
\label{LapH2}
\end{equation}
\begin{equation}
\frac{1}{\rho}\frac{\partial H_1}{\partial {\rho}} +  \frac{\partial^2 H_1}{\partial {{\rho}^2}} +\frac{1}{V(r)} \left( \frac{2}{r}\frac{\partial H_1}{\partial {r}} +  \frac{\partial^2 H_1}{\partial {{r}^2}} \right)=-H_2 \left(  \frac{\partial^2 H_1}{\partial {x_1}^2} + \frac{\partial^2 H_1}{\partial {x_2}^2} \right).
\label{LapH1}
\end{equation}
The solutions to equation (\ref{LapH1}) for the first membrane metric function $H_1$ depend on the solutions to equation (\ref{LapH2}) for the second membrane metric function $H_2$. We can find the solutions to the first differential equation \eqref{LapH2} by separating the coordinates as $H_2=\mathcal{Q}_2F_1(\rho)F_2(r)$ where $\mathcal{Q}_2$ is the charge of 
M2 brane. This separation of variables gives an exact solution for $H_2$ 
in the near core region of second M2 brane. 
The equation \eqref{LapH2} reduces to two ordinary decoupled differential equations for $F_1(\rho)$ and $F_2(r)$ that are given by
\begin{equation}
\frac{d^2 F_1(\rho)}{d\rho^2}=c^2 F_1(\rho) -\frac{1}{\rho}\frac{d F_1(\rho)}{d\rho},
\label{odeF1C}
\end{equation}
\begin{equation}
\frac{d^2 F_2(r)}{dr^2}=-c^2 V(r) F_2(r) -\frac{2}{r}\frac{d F_2(r)}{dr},
\label{odeF2C}
\end{equation}
where $c$ is the separation constant. The solutions to \eqref{odeF1C} are given by
\begin{equation}
F_1(\rho) = C_1 I_0(c\rho)+C_2 K_0(c\rho),
\label{odeF1CS}
\end{equation}
in terms of modified Bessel functions $I_0$ and $K_0$ and $C_1$, $C_2$ are two constants. To have a finite solution far away from the branes, we have to choose the constant $C_1=0$. Moreover, the solutions to (\ref{odeF2C}) are given by
\begin{equation}
F_2(r)= e^{-icr}K_M(1+\frac{1}{2}icn,2,2icr) \left( B_1 + B_2 \int \frac{dr}{r^2 e^{-2icr} {{K_M}(1+\frac{1}{2}icn,2,2icr)}^2} \right),
\label{odeF2CS}
\end{equation}
where $K_M$ stands for the Kummer function of type $M$ and $B_1$, $B_2$ are two constants. In what follows in the present section, we show the first and second term of (\ref{odeF2CS}) by $\mathcal{F}_1(r)$ and $\mathcal{F}_2(r)$ respectively. Figure \ref{fig:figure1} shows the behaviour of solution $\mathcal{F}_1(r)$, where the solution vanishes at infinity. 
The near horizon limit of $\mathcal{F}_2(r)$ (equation (\ref{limit2})) requires that we choose the constant $B_2=0$ in (\ref{odeF2CS}).  
The general solution for the metric function $H_2(\rho,r)$ is a superposition of different solutions given by (\ref{odeF1CS}) and (\ref{odeF2CS}) for different values of $c$,
\begin{equation}
H_2(\rho,r)=\mathcal{Q}_2 \int_0^{+\infty} g(c) F_1(\rho)F_2(r) dc, 
\label{MH2CS}
\end{equation}
where $g(c)$ is an arbitrary function of the separation constant $c$.
To determine and fix the function $g(c)$, we consider the limits
\begin{subequations}\label{limitFlat}
\begin{align} 
& r \rightarrow 0 \label{limitFlat1}, \\
&n \rightarrow +\infty \label{limitFlat2}, \\
&nr=\text{fixed}, \label{limitFlat3} 
\end{align}
\end{subequations}
in which the transverse space to M2 branes approaches to $E^4 \bigotimes D^2$
\begin{equation}
ds_{6}^{2}(r,\rho)=dz^2+z^2 {d\Omega}_3+d{\rho}^2+ {\rho}^2{d\eta}^2,\label{sixdimlimits}
\end{equation}
where $z=2\sqrt{nr}$. To obtain (\ref{sixdimlimits}),  
we consider $V(r) \approx \frac{n}{r}$ and the coordinate $\psi$ is redefined to $\tilde{\psi}$ by ${\psi}=2n\tilde{\psi}$.
In the limits (\ref{limitFlat}) where the transverse space to M2 branes is flat, the metric function $H_2(\rho,r)$ must approach to $\frac{\mathcal{Q}_2}{(z^2+\rho^2)^2}$. The proper choice of $g(c)$ in the general solution (\ref{MH2CS}) leads to the required behaviour of the metric function $H_2(\rho,r)$ in the limits (\ref{limitFlat}). In the limits (\ref{limitFlat}), the differential equation (\ref{odeF2C}) for $F_2(r)$
can be simplified to 
\begin{equation}
\frac{d^2 F_2(r)}{dr^2}=-\frac{n c^2}{r} F_2(r) -\frac{2}{r}\frac{d F_2(r)}{dr}.
\label{odeF2CAPP}
\end{equation}
The solutions to (\ref{odeF2CAPP}) are given by 
\begin{equation}
F_2(r)=E_1\frac{1}{\sqrt{r}} J_1(2c\sqrt{nr})+E_2\frac{1}{\sqrt{r}}Y_1(2c\sqrt{nr}),
\label{odeF2CAS}
\end{equation}
in terms of Bessel functions where $E_1$ and $E_2$ are two constants. Comparing \eqref{odeF2CS} with \eqref{odeF2CAS}, one can show  
\begin{eqnarray}
\mathop {\lim }\limits_{{\scriptstyle r \to 0 \hfill \atop \scriptstyle n \to  +\infty \hfill} \atop \scriptstyle nr=\text{fixed} } \mathcal{F}_1 &\approx&  \frac{J_1(2c\sqrt{nr})}{c\sqrt{nr}}, \label{limit1} \\
\mathop {\lim }\limits_{{\scriptstyle r \to 0 \hfill \atop \scriptstyle n \to  +\infty \hfill} \atop \scriptstyle nr=\text{fixed} } \mathcal{F}_2 &\approx&  \frac{c\pi\sqrt{n}Y_1(2c\sqrt{nr})}{\sqrt{r}}. \label{limit2}
\end{eqnarray}

Hence the metric function \eqref{MH2CS} in the limits (\ref{limitFlat}) becomes 
\begin{eqnarray}
H_2(\rho,r)&\approx& \mathcal{Q}_2 \int_0^{+\infty} g(c) F_1(\rho){\bigg|}_{C_1=0 \hfill \atop C_2=1 \hfill }F_2(r) {\bigg|}_{B_1=1 \hfill \atop B_2=0 \hfill } dc \label{limitH1}\\
&=&\frac{2\mathcal{Q}_2}{z}  \int_0^{+\infty} g(c) K_0(c\rho) J_1(cz) \frac{dc}{c} =\frac{\mathcal{Q}_2 }{(z^2+\rho^2)^2}\bigg|_{z=2\sqrt{nr}}. 
\label{limitH2} 
\end{eqnarray}
We choose $B_2=0$ in integrand of (\ref{limitH1}) since in the limits (\ref{limitFlat}), the radial function $\mathcal{F}_2$ diverges.
The solution to integral equation (\ref{limitH2}) for $g(c)$ is given by $g(c)=\displaystyle\frac{c^3}{4}$. So the general  
solution for the second membrane metric function $H_2(\rho,r)$ reads as
\begin{equation}
H_2(\rho,r)=\frac{\mathcal{Q}_2}{4} \int_0^{+\infty} c^3  K_0(c\rho) e^{-icr}K_M(1+\frac{1}{2}icn,2,2icr) dc. 
\label{MH2CSTAUB}
\end{equation}
Furnished with the general solution for $H_2(\rho,r)$, we can find the general solution to the second differential equation \eqref{LapH1} for the first membrane metric function $H_1(x_1,x_2,\rho,r)$. The differential equation (\ref{LapH1}) is separable if we separate the coordinates by
\begin{equation}
H_1(x_1,x_2,\rho,r)=1+\mathcal{Q}_1 (x_1^2+x_2^2)+F_1(\rho)h(r).
\label{H1MetricFunction}
\end{equation}
where $\mathcal{Q}_1$ is the charge of first M2 brane. Substituting (\ref{H1MetricFunction}) in (\ref{LapH1}) and using equation (\ref{odeF1C}) for $F_1(\rho)$, we get the differential equation for $h(r)$
\begin{equation}
\frac{d^2h(r)}{dr^2}+\frac{2}{r}\frac{dh(r)}{dr}+c^2 V(r) h(r)=\mathcal{G}(r),
\label{SecondH}
\end{equation}
where $\mathcal{G}(r)=-4\mathcal{Q}_1 \mathcal{Q}_2 F_2(r)V(r)$. The solution to (\ref{SecondH}) is 
\begin{equation}
h(r)=-\mathcal{F}_1(r) \int{\frac{\mathcal{G}(r) \mathcal{F}_2(r)dr}{\mathcal{W}(\mathcal{F}_1(r),\mathcal{F}_2(r))}}+\mathcal{F}_2(r) \int{\frac{\mathcal{G}(r) \mathcal{F}_1(r)dr}{\mathcal{W}(\mathcal{F}_1(r),\mathcal{F}_2(r))}},
\label{hrps}
\end{equation}
where $\mathcal{W}(\mathcal{F}_1(r),\mathcal{F}_2(r))$ is the Wronskian of $\mathcal{F}_1(r)$ and $\mathcal{F}_2(r)$. Hence, the general solution for the metric function of first membrane $H_1(x_1,x_2,\rho,r)$ is a superposition of solutions (\ref{odeF1CS}) and (\ref{hrps}) of the form
\begin{equation}
H_1(x_1,x_2,\rho,r)=1+\mathcal{Q}_1 (x_1^2+x_2^2)+\int_{0}^{\infty} \tilde{g}(c) F_1(\rho)h(r) dc.
\label{H1TAUB}
\end{equation}
The weight function $\tilde{g}(c)$ in (\ref{H1TAUB}) can be determined by comparing the general solution (\ref{H1TAUB}) 
to the well known solution of M2 brane in the limits (\ref{limitFlat}) where the six dimensional flat transverse space to membranes has the line element (\ref{sixdimlimits}).
In the limits (\ref{limitFlat}), the source term in the inhomogeneous differential equation (\ref{SecondH}) approaches to
\begin{equation}
\mathcal{G}(r) \approx -4\mathcal{Q}_1 \mathcal{Q}_2 \frac{nJ_1(2c\sqrt{nr})}{cr\sqrt{nr}}.
\end{equation}
So, we find that in the limits (\ref{limitFlat}), the solution to inhomogeneous equation (\ref{SecondH}) for $h(r)$ is given by
\begin{equation}
h(r)=\frac{4\mathcal{Q}_1 \mathcal{Q}_2}{c^2} J_0(2c\sqrt{nr}).
\label{Zr.2}
\end{equation}
Moreover, a tedious calculation shows that the special solution (\ref{Zr.2}) also can be obtained directly from the solution (\ref{hrps}) in appropriate limits (\ref{limitFlat}).
The dependence of first membrane metric function $H_1$ on $r$ and $\rho$ in the background of transverse space to M2 branes is given by $\frac{\mathcal{Q}_1 \mathcal{Q}_2 }{z^2+\rho^2}$ \cite{Youam}.
Hence, we get 
\begin{equation}
4\mathcal{Q}_1 \mathcal{Q}_2  \int_{0}^{\infty} \frac{\tilde{g}(c) }{c^2} K_0(
c\rho) J_0(2c\sqrt{nr}) dc = \frac{\mathcal{Q}_1 \mathcal{Q}_2 }{z^2+\rho^2}\bigg|_{z=2\sqrt{nr}},
\label{intforg}
\end{equation}
as an integral equation for the function  $\tilde{g}(c)$. The solution to equation (\ref{intforg}) is $\tilde{g}(c)=\displaystyle\frac{c^3}{4}$ and so we get the general solution for the M2 brane function $H_1(x_1,x_2,\rho,r)$  
\begin{equation}
H_1(x_1,x_2,\rho,r)
=1+\mathcal{Q}_1 (x_1^2+x_2^2)+\frac{1}{4}\int_{0}^{\infty} c K_0(c\rho) h(r) dc,
\label{H1TAUBF}
\end{equation}
where $h(r)$ is given by (\ref{hrps}). 

In addition to exact membrane solutions (\ref{MH2CSTAUB}) and (\ref{H1TAUBF}),  we can obtain a new set of solutions for the membrane metric 
functions by analytically continuing the separation constant $c$ that appears in separated differential equations (\ref{odeF1C}) and (\ref{odeF2C}) to $i\tilde c$.
The differential equation \eqref{odeF1C} changes to 
\begin{equation}
\frac{d^2 \tilde{F}_1(\rho)}{d\rho^2}=-\tilde c^2 \tilde{F}_1(\rho) -\frac{1}{\rho}\frac{d \tilde{F}_1(\rho)}{d\rho},
\label{odeF1IC}
\end{equation}
where the solutions are given by
\begin{equation}
\tilde{F}_1(\rho) = \tilde{C}_1 Y_0(\tilde c\rho)+\tilde{C}_2 J_0(\tilde c\rho).
\label{odeF1ICS}
\end{equation}
in terms of Bessel functions. 

\begin{figure}[H]
\centering
 \includegraphics[width=0.5\textwidth]{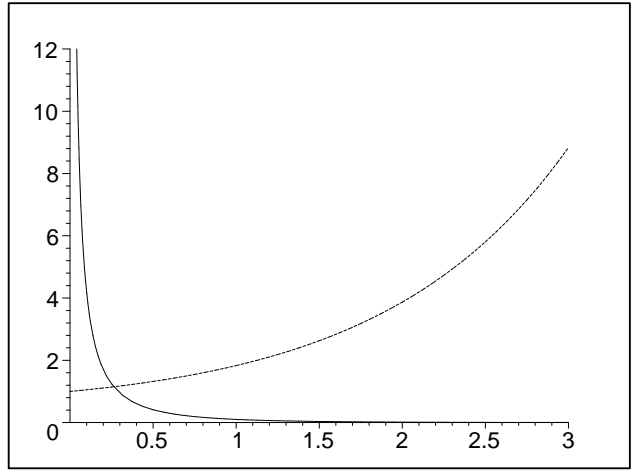}
 \caption{Two independent solutions $\mathcal{\hat{F}}_1(r)$ (dashed) and $\mathcal{\hat{F}}_2(r)$ (solid) in equation (\ref{odeF2ICS}).
}
 \label{fig:figure2}
\end{figure}

The other differential equation \eqref{odeF2C} takes the form
\begin{equation}
\frac{d^2 \tilde{F}_2(r)}{dr^2}=\tilde c^2 V(r) \tilde{F}_2(r) -\frac{2}{r}\frac{d \tilde{F}_2(r)}{dr}.
\label{odeF2IC}
\end{equation}
The solutions to (\ref{odeF2IC}) are
\begin{equation}
\tilde{F}_2(r) = \tilde{B}_1 e^{-\tilde cr}K_M(1+\frac{1}{2}\tilde cn,2,2\tilde cr)+\tilde{B}_2 e^{-\tilde cr}K_U(1+\frac{1}{2}\tilde cn,2,2\tilde cr),
\label{odeF2ICS}
\end{equation}
where $K_M$ and $K_U$ are the Kummer functions of type $M$ and $U$ respectively  (figure \ref{fig:figure2}) and $\tilde{B}_1$, $\tilde{B}_2$ are two constants. 
We show the first and second term of (\ref{odeF2ICS})
by $\mathcal{\hat{F}}_1(r)$ and $\mathcal{\hat{F}}_2(r)$.
The only acceptable solution at infinity is $\mathcal{\hat{F}}_2(r)$ and so we may write the general solution for the metric function $\tilde{H}_2$ as
\begin{equation}
\tilde{H}_2(\rho,r)
=\mathcal{Q}_2 \int_0^{+\infty} \tilde{g}_2(\tilde c) e^{-\tilde cr} K_U(1+\frac{\tilde cn}{2},2,2\tilde cr)J_0(\tilde c\rho) d\tilde c. 
\label{MH2ICS}
\end{equation}

To determine the weight function $ \tilde{g}_2(\tilde c)$, we consider the limits \eqref{limitFlat}, in which one can show 
\begin{equation}
\mathop {\lim }\limits_{{\scriptstyle r \to 0 \hfill \atop \scriptstyle n \to  +\infty \hfill} \atop \scriptstyle nr=\text{fixed} } e^{-\tilde cr} K_U(1+\frac{\tilde cn}{2},2,2\tilde cr) \approx  \frac{2K_1(2\tilde c\sqrt{nr})}{\tilde c\sqrt{nr}\Gamma(\frac{\tilde \tilde cn}{2})}. \\
\label{AppIC1}
\end{equation}
Consequently, we get the integral equation for $ \tilde{g}_2(\tilde c)$ as
\begin{equation}
\mathcal{Q}_2 \int_0^{+\infty} \tilde{g}_2(\tilde c) \frac{K_1(2\tilde c\sqrt{nr})}{\tilde c\sqrt{nr}\Gamma(\frac{\tilde \tilde cn}{2})} J_0(\tilde c\rho) d\tilde c=\frac{\mathcal{Q}_2}{2(z^2+\rho^2)^2}\bigg|_{z=2\sqrt{nr}}.
\label{MH2ICSFLAT}
\end{equation}
We find the solution to integral equation (\ref{MH2ICSFLAT}) for $\tilde{g}_2(\tilde c)$ is given by
\begin{equation}
\tilde{g}_2(\tilde c)=\displaystyle\frac{c^3}{8}\Gamma(\frac{\tilde cn}{2}),
\end{equation}
and so the general solution for the metric function $\tilde{H}_2$ is  
\begin{equation}
\tilde{H}_2(\rho,r)
=\mathcal{Q}_2 \int_0^{+\infty} \frac{\tilde c^3}{8} \Gamma(\frac{\tilde cn}{2}) e^{-\tilde cr} K_U(1+\frac{\tilde cn}{2},2,2\tilde cr)J_0(\tilde c\rho) d\tilde c. 
\label{MH2ICS2}
\end{equation}
Furnished by the new solution (\ref{MH2ICS2}) for $\tilde{H}_2$, we separate the coordinates in the first membrane metric function by 
\begin{equation}
\tilde H_1(x_1,x_2,\rho,r)=1+\mathcal{Q}_1 (x_1^2+x_2^2)+\tilde F_1(\rho)\tilde h(r),\label{H2tildesep}
\end{equation}
to separate the differential equation (\ref{LapH1}). Plugging (\ref{H2tildesep}) in 
(\ref{LapH1}) and using equation (\ref{odeF1ICS}) for $\tilde F_1(\rho)$, we find the following equation for $\tilde h(r)$
\begin{equation}
\frac{d^2\tilde{h}(r)}{dr^2}+\frac{2}{r}\frac{d\tilde{h}(r)}{dr}-\tilde c^2 (1+\frac{n}{r}) \tilde{h}(r)=\tilde{\mathcal{G}}(r),
\label{SecondHH}
\end{equation}
where $\tilde{\mathcal{G}}(r)=-4\mathcal{Q}_1 \mathcal{Q}_2  (1+\frac{n}{r})  e^{-\tilde cr} K_U(1+\frac{1}{2}\tilde cn,2,2\tilde cr)$. The solution to (\ref{SecondHH}) is 
\begin{equation}
\tilde h(r)=-\hat {\mathcal{F}}_1(r) \int{\frac{\tilde {\mathcal{G}}(r) \hat{\mathcal{F}}_2(r)dr}{\mathcal{W}(\hat{\mathcal{F}}_1(r),\hat{\mathcal{F}}_2(r))}}+\hat{\mathcal{F}}_2(r) \int{\frac{\tilde {\mathcal{G}}(r) \hat{\mathcal{F}}_1(r)dr}{\mathcal{W}(\hat{\mathcal{F}}_1(r),\hat{\mathcal{F}}_2(r))}},
\label{hrpsIC}
\end{equation}
where $\mathcal{W}(\hat{\mathcal{F}}_1(r),\hat{\mathcal{F}}_2(r))$ is the Wronskian of $\hat{\mathcal{F}}_1(r)$ and $\hat{\mathcal{F}}_1(r)$. 
So, we can write the general solution for the metric function $\tilde{H}_1$ as  
\begin{equation}
\tilde{H}_1(x_1,x_2,\rho,r)=1+\mathcal{Q}_1 (x_1^2+x_2^2)+\int_{0}^{\infty} \tilde{g}_1(\tilde c) J_0(\tilde c\rho)\tilde{h}(r) d\tilde c,
\label{H1TAUBIC}
\end{equation}
where $\tilde{h}(r)$ is given by \eqref{hrpsIC}. 
To determine and fix the weight function $\tilde{g}_1(\tilde c)$, we compare the solution (\ref{H1TAUBIC}) to the known solution of membrane function in the limits 
(\ref{limitFlat}), where the transverse space to membranes becomes flat. In the limits (\ref{limitFlat}), the 
differential equation (\ref{SecondHH}) reduces to
\begin{equation}
\frac{d^2\tilde{h}(r)}{dr^2}+\frac{2}{r}\frac{d\tilde{h}(r)}{dr}-\frac{\tilde c^2n}{r}\tilde{h}(r)=-8n\mathcal{Q}_1 \mathcal{Q}_2  \frac{K_1(2\tilde c\sqrt{rn})}{r\tilde c\sqrt{nr}\Gamma(\frac{\tilde cn}{2})}.
\label{SecondHHFLAT}
\end{equation}
We find that the solution to equation (\ref{SecondHHFLAT}) is given by 
\begin{equation}
\tilde{h}(r)=\frac{8\mathcal{Q}_1\mathcal{Q}_2}{c^2\Gamma(\frac{\tilde cn}{2})}K_0(2\tilde c\sqrt{nr}).
\label{hICFLAT}
\end{equation}
Comparing the equation (\ref{H1TAUBIC}) in which $\tilde h(r)$ is given by (\ref{hICFLAT}) to the solution of membrane function where the transverse space is flat yields an integral equation for $\tilde{g}_1(\tilde c)$.
The solution to the integral equation is given by $\tilde{g}_1(\tilde c)=\frac{\tilde c^3\Gamma(\frac{\tilde cn}{2})}{8}$ and so, the metric function for the first M2 brane takes
the form 
\begin{equation}
\tilde{H}_1(x_1,x_2,\rho,r)
=1+\mathcal{Q}_1 (x_1^2+x_2^2)+\int_{0}^{\infty} \frac{\tilde c^2}{4} J_0(\tilde c\rho) \tilde{h}(r) d\tilde c. \label{HHtilde}
\end{equation}

Dimensional reduction of all two-membrane solutions in this section yields  
the field content and metric of type IIA string theory. The dimensional reduction along the coordinate $\psi$ of (\ref{H1H2H}) gives the NSNS dilaton and the RR one-form 
by
$\Phi=\frac{3}{4} \ln(\frac{{H_1}^{\frac{1}{3}}{H_2}^{\frac{1}{3}}}{V})$ and $C_{\phi}=n \cos(\theta)$. 
The antisymmetric NSNS two-form is zero and the only non-zero components of RR three-form are 
$A_{tx_1x_2}=\frac{1}{2H_2}$ and $
A_{ty_1y_2}=\frac{1}{2H_1}$.
The ten-dimensional metric read as
\begin{eqnarray}
ds^2&=&-H_1^{-\frac{1}{2}}H_2^{-\frac{1}{2}}V^{-\frac{1}{2}} dt^2 + H_1^{\frac{1}{2}}H_2^{-\frac{1}{2}}V^{-\frac{1}{2}}(dx_1^2+dx_2^2)+ H_1^{-\frac{1}{2}}H_2^{\frac{1}{2}}V^{-\frac{1}{2}}(dy_1^2+dy_2^2)+\nonumber\\
&+& H_1^{\frac{1}{2}}H_2^{\frac{1}{2}}V^{-\frac{1}{2}}(d{\rho}^2+{\rho}^2 {d\alpha}^2)+ H_1^{\frac{1}{2}}H_2^{\frac{1}{2}}V^{\frac{1}{2}}\left[dr^2+r^2 \left( {d\theta}^2+ \sin^2(\theta) {d\phi}^2 \right) \right],
\label{metri}
\end{eqnarray}
that describes a system of three D branes.
We have explicitly checked the solution (\ref{metri}) along with the dilaton, RR one-form and RR three form exactly satisfy all the supergravity equations of motion in ten dimensions. For the second set of solutions (\ref{MH2ICS2}) and (\ref{HHtilde}), $H_{1}$ and $H_2$ should be replaced by $\tilde{H}_1$ and $\tilde{H}_2$ in the ten-dimensional NSNS and RR fields and the metric (\ref{metri}).
\section{Embedding Bianchi IX space in a system of two M2 branes}
\label{sec:Bianchi}
In this section, we consider the triaxial Bianchi IX space (\ref{BIX}) as a part of transverse space to two M2 branes. Requiring that the metric \eqref{ds11genM2} and the four-form field strength (\ref{4Field}) satisfy the equations of motion (\ref{GminGG}), (\ref{dF}) gives the following two coupled differential equations for the metric functions $H_1$ and $H_2$

\begin{eqnarray}
&&\left\{ 2\frac{\displaystyle\frac{df(r)}{dr}}{\sqrt{f(r)}} +\sqrt{f(r)}\bigg[ -\sum_{i=1}^{3} \frac{\displaystyle\frac{df_i(r)}{dr}}{f_i(r)}+\frac{6}{r} \bigg] \right\} \frac{\partial H_2}{\partial r}+2\sqrt{f(r)}\frac{\partial^2 H_2}{\partial r^2}+2\left[ \frac{\partial^2 H_2}{\partial {\rho}^2}+\frac{1}{\rho}\frac{\partial H_2}{\partial {\rho}} \right]=0, \nonumber \\
\label{Biaeq2}
\end{eqnarray}
\begin{eqnarray}
\left\{ 2\frac{\displaystyle\frac{df(r)}{dr}}{\sqrt{f(r)}} +\sqrt{f(r)}\bigg[ -\sum_{i=1}^{3} \frac{\displaystyle\frac{df_i(r)}{dr}}{f_i(r)}+\frac{6}{r} \bigg] \right\} \frac{\partial H_1}{\partial r}&+&2\sqrt{f(r)}\frac{\partial^2 H_1}{\partial r^2}+2\left[ \frac{\partial^2 H_1}{\partial {\rho}^2}+\frac{1}{\rho}\frac{\partial H_1}{\partial {\rho}} \right] \nonumber \\
&=&-2H_2\left[ \frac{\partial^2 H_1}{\partial x_1^2}+\frac{\partial^2 H_1}{\partial x_2^2} \right]. 
\label{Biaeq1}
\end{eqnarray}
We choose $k=1$ which means $a_1=0$ and $a_2=a_3=2b$ that we show simply by $a$. The differential equations (\ref{Biaeq2}) and (\ref{Biaeq1}) then reduce to
\begin{eqnarray}
(\frac{a^4}{r^5}+\frac{3}{r}) \frac{\partial }{\partial r} H_2(\rho,r)+(1-\frac{a^4}{r^4}) \frac{\partial^2 }{\partial r^2} H_2(\rho,r)+ \frac{1}{\rho}\frac{\partial }{\partial \rho} H_2(\rho,r) +\frac{\partial^2 }{\partial \rho^2} H_2(\rho,r)&=&0,\nonumber \\ && \label{PdeP.2}\\
(\frac{a^4}{r^5}+\frac{3}{r}) \frac{\partial }{\partial r} H_1(x_1,x_2,\rho,r)+(1-\frac{a^4}{r^4}) \frac{\partial^2 }{\partial r^2} H_1(x_1,x_2,\rho,r)+ \frac{1}{\rho}\frac{\partial }{\partial \rho} H_1(x_1,x_2,\rho,r)&+& \nonumber \\
+\frac{\partial^2 }{\partial \rho^2} H_1(x_1,x_2,\rho,r)+ H_2(\rho,r)\left(\frac{\partial^2 }{\partial x_1^2} H_1(x_1,x_2,\rho,r)+\frac{\partial^2 }{\partial x_2^2} H_1(x_1,x_2,\rho,r) \right)&=&0. \nonumber \\ && \label{PdeP.1}
\end{eqnarray}

\begin{figure}[H]
\centering
 \includegraphics[width=0.5\textwidth]{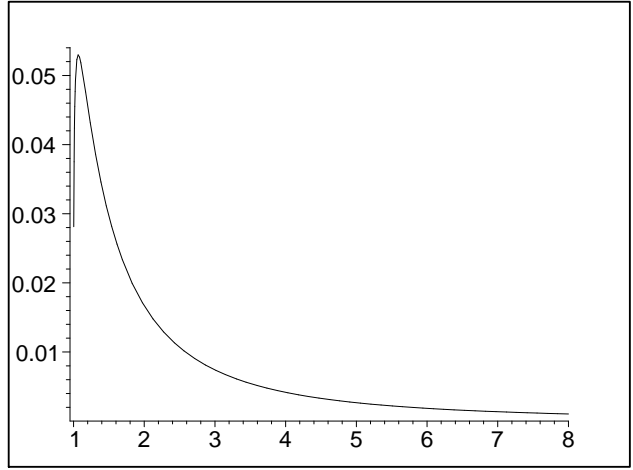}
 \caption{The difference between $\frac{\cosh(t)}{\sinh^3(t)}$ and $\frac{1}{ t^3}$ as function of $\frac{r}{a}$.}
 \label{diffplot}
\end{figure}

\begin{figure}[H]
\centering
\includegraphics[width=0.5\textwidth]{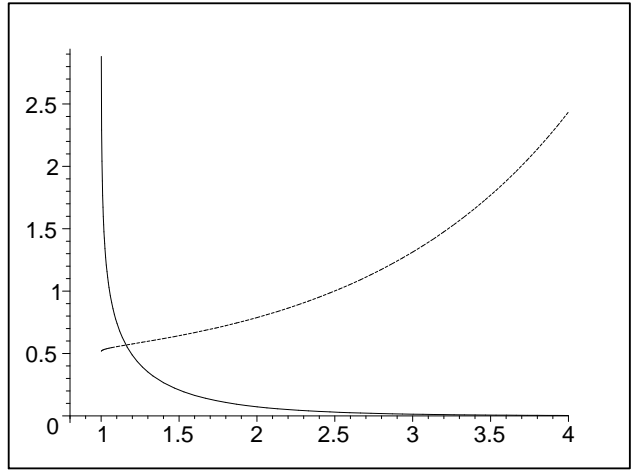}
 \caption{$h_1$ and $h_2$ solutions in (\ref{twoS}) as function of $\frac{r}{a}$ where we set $a=1$. }
 \label{fig:figure16}
\end{figure}

\begin{figure}[H]
\centering
 \includegraphics[width=0.5\textwidth]{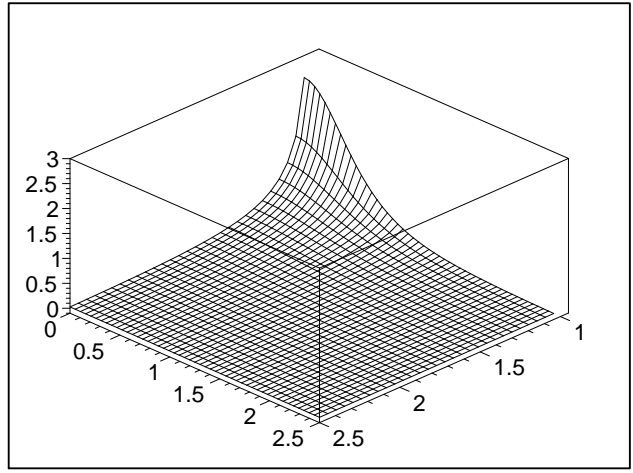}
 \caption{The behaviour of $H_2$ (\ref{H2TSOL}) versus $0 \leq \rho \leq 2.5$ and $1 \leq r \leq 2.5$, where we set $a=\mathcal{Q}_2=1$.}
 \label{fig:figure43}
\end{figure}

To find the solutions to differential equation (\ref{PdeP.2}), we separate the coordinates as $H_2(\rho,r)=f(\rho)h(r)$.  
We get two separated ordinary differential equations for $f(\rho)$ and $h(r)$ that are given by
\begin{eqnarray}
\frac{1}{\rho f}\frac{df}{d\rho}+\frac{1}{f}\frac{d^2f}{d\rho^2}+c^2&=&0, \label{pde.1} \\
\frac{\left( 3+\frac{a^4}{r^4} \right)}{rh}\frac{dh}{dr}+\frac{\left( 1-\frac{a^4}{r^4} \right)}{rh}\frac{d^2h}{dr^2}-c^2&=&0,  \label{pde.2}  
\end{eqnarray}
where $c$ is the separation constant.
The solutions to \eqref{pde.1} are 
\begin{equation}
f(\rho)=C_1 J_0(c\rho)+C_2 Y_0(c\rho),\label{gofrho}
\end{equation}
where $C_1$ and $C_2$ are two constants. To find the solutions to (\ref{pde.2}), we define the new positive variable $t$ by $r=\frac{ a}{\sqrt{\tanh(t)}}$. In terms of variable $t$, the differential equation \eqref{pde.2} becomes
\begin{equation}
\frac{d^2h(t)}{dt^2}-\frac{c^2a^2}{4}\frac{\cosh(t)}{\sinh^3(t)}h(t)=0.
\label{eqcosh}
\end{equation}
Although it is very unlikely to find any exact analytic solutions to equation (\ref{eqcosh}), however we can approximate $\frac{\cosh(t)}{\sinh^3(t)}$ by $\frac{1}{t^3}$ to find the approximate analytic solutions. Figure \ref{diffplot} shows the difference between the actual function and the approximation, $\delta=\frac{1}{ t^3}-\frac{\cosh(t)}{sinh^3(t)}$, as a function of $\frac{r}{a}$. The maximum difference $\delta_{max}=0.053$ occurs at $\frac{r}{a}\simeq 1.067$.  
The analytic solutions to equation (\ref{eqcosh}) are given by
\begin{equation}
h(r)= 
D_1 \sqrt{\tanh^{-1}(\frac{a^2}{r^2})} I_1(\frac{ca}{\sqrt{\tanh^{-1}(\frac{a^2}{r^2})}}) + D_2 \sqrt{\tanh^{-1}(\frac{a^2}{r^2})} K_1(\frac{ca}{\sqrt{\tanh^{-1}(\frac{a^2}{r^2})}}),
\label{twoS}
\end{equation}
in terms of modified Bessel functions where $D_1$ and $D_2$ are constants. Figure \ref{fig:figure16} shows the typical behaviour of solutions (\ref{twoS}) in terms of $\frac{r}{a}$ where $h_1$ and $h_2$ are the first and second terms in (\ref{twoS}) respectively.
We may write the general solution for the metric function $H_2$ as a superposition of the different solutions for $f(\rho)$ and $g(r)$ with different values of $c$
\begin{equation}
H_2(\rho,r)=\mathcal{Q}_2 \int_0^{+\infty} g_2(c) h_2(r) J_0(c\rho) dc, 
\label{MH2CSBIX}
\end{equation}
where we choose the constants $C_2=D_1=0$ in (\ref{gofrho}) and (\ref{twoS}) and $g_2(c)$ is the weight function. The weight function can be fixed by comparing the equation (\ref{MH2CSBIX}) in the limit $t \rightarrow 0$, where the transverse space to membranes is flat, to the metric function $H_2=\frac{\mathcal{Q}_2}{(\rho^2+r^2)^2}$ in the asymptotic flat region.
In the asymptotic flat limit, one can set $\tanh(t) \approx t$ and so $ r\approx \frac{a}{\sqrt{t}}$. Hence the solutions (\ref{twoS}) reduce to
\begin{equation}
h(r)=D_2 \frac{a}{r} K_1(cr),
\end{equation}
and so the metric function $H_2$ (\ref{MH2CSBIX}) in asymptotic region must satisfy
\begin{equation}
a\int_0^{+\infty}  g_2(c) \frac{K_1(cr)}{r} J_0(c\rho) dc=\frac{1}{(\rho^2+r^2)^2}.\label{inteqforH2}
\end{equation}
The solution to integral equation (\ref{inteqforH2}) is $ g_2(c)=\frac{c^2}{2a}$ and so we get the general form of the metric function $H_2$ 
\begin{equation}
H_2(\rho,r)=\frac{\mathcal{Q}_2}{2a}\sqrt{\tanh^{-1}(\frac{a^2}{r^2})}\int_0^{+\infty} {c^2}J_0(c\rho) { K_1(\frac{ca}{\sqrt{\tanh^{-1}(\frac{a^2}{r^2})}})} dc. 
\label{H2TSOL}
\end{equation}
Quite interestingly, the integral in (\ref{H2TSOL}) can be done analytically and we get $H_2(\rho,r)=
\mathcal{Q}_2\frac{\left\{ \tanh^{-1}(\frac{a^2}{r^2}) \right\}^2}{\left\{ \rho^2{\tanh^{-1}(\frac{a^2}{r^2})}+a^2 \right\} ^2}.
$
Figure \ref{fig:figure43} shows how $H_2$ varies versus $\rho$ and $r$ where we set $a=\mathcal{Q}_2=1$.

To find the solutions to \eqref{PdeP.1} for $H_1(x_1,x_2,\rho,r)$, we consider 
\begin{equation}
H_1(x_1,x_2,\rho,r)=1+\mathcal{Q}_1(x_1^2+x_2^2)+f(\rho)k(r), \label{H1sep}  
\end{equation}
where $f(\rho)$ satisfies \eqref{pde.1}. Upon substituting (\ref{H1sep}) in equation (\ref{PdeP.1}), we find the differential equation for $k(r)$ that is given by 
\begin{equation}
\left(1-\frac{a^4}{r^4}  \right) \frac{d^2}{dr^2}k(r) + \frac{1}{r}\left( 3+\frac{a^4}{r^4} \right)\frac{d}{dr}k(r)-c^2 k(r)=-4\mathcal{Q}_1\mathcal{Q}_2h_2(r),
\label{odehrc}
\end{equation}
or in terms of new positive
variable $t$ as
\begin{equation}
\frac{d^2}{dt^2} k(t)-\frac{c^2 a^2}{4} \frac{\cosh(t) }{\sinh^3(t)}k(t)=-a^2\mathcal{Q}_1\mathcal{Q}_2h_2(t)\frac{\cosh(t) }{\sinh^3(t)}.
\label{odeTtc}
\end{equation}
%
To find the approximate analytic solutions to equation (\ref{odeTtc}), we approximate $\frac{\cosh(t) }{\sinh^3(t)}$ by $\frac{1}{t^3}$ in equation \eqref{odeTtc} which yields
\begin{equation}
t^3\frac{d^2}{dt^2} k(t)-\frac{c^2 a^2}{4} k(t)=-a^2\mathcal{Q}_1\mathcal{Q}_2 \sqrt{t}K_1(\frac{ca}{\sqrt{t}}). 
\label{odeTtcS}
\end{equation}
The solution to (\ref{odeTtcS}) is given by
\begin{equation}
k(t)=\frac{2a\mathcal{Q}_1\mathcal{Q}_2}{c} K_0(\frac{ca}{\sqrt{t}}),
\label{odeTtcSS}
\end{equation}
and so, 
the general approximate analytical solution for the metric function $H_1$ is 
\begin{equation}
\begin{split}
H_1(x_1,x_2,\rho,r)&=1+\mathcal{Q}_1(x_1^2+x_2^2)+\int_0^{\infty}g_1(c) f(\rho)k(r)dc \\
&=1+\mathcal{Q}_1(x_1^2+x_2^2)+\frac{2a\mathcal{Q}_1\mathcal{Q}_2}{c}\int_0^{\infty}g_1(c) J_0(c\rho)  K_0(\frac{ca}{\sqrt{\tanh^{-1}\left(\frac{a^2}{r^2}\right)}}) dc.
\label{H1TSOL1}
\end{split}
\end{equation}
In (\ref{H1TSOL1}), the weight function $g_1(c)$ must be chosen such that in asymptotic flat region, the solution (\ref{H1TSOL1}) approaches to $1+\mathcal{Q}_1(x_1^2+x_2^2)+\frac{\mathcal{Q}_1\mathcal{Q}_2}{\rho^2+r^2}$.
One can find the appropriate solution for $g_1(c)$ is $g_1(c)=\frac{c^2}{2a}$ and finally we find 
\begin{equation}
H_1(x_1,x_2,\rho,r)=1+\mathcal{Q}_1(x_1^2+x_2^2)+\mathcal{Q}_1\mathcal{Q}_2\frac{\tanh^{-1} \left( \frac{a^2}{r^2} \right)}{a^2 + \rho^2 \tanh^{-1} \left( \frac{a^2}{r^2} \right) }.
\label{H1sol}
\end{equation}
Although it seems we may obtain a second set of solutions for the metric functions by analytically continuing the separation constant $c$ to $i\tilde c$, however the calculation shows that the second set of solutions are exactly the same as the solutions (\ref{H2TSOL}) and (\ref{H1sol}).
\section{Supersymmetry}
\label{sec:Supersymmetry}

The number of preserved supersymmetry for any solution of supergravity in eleven dimensions is determined by the number of solutions to the Killing spinor equation. The Killing spinor equation is given by \cite{gauntlett}
\begin{equation}
\partial_{M} \varepsilon+\frac{1}{4}\omega_{abM}\Gamma^{ab} \varepsilon+\frac{1%
} {144}\Gamma_{M}^{\phantom{M}NPQR}F_{NPQR} \varepsilon-\frac{1}{18}\Gamma
^{PQR}F_{MPQR} \varepsilon=0,  \label{killingspinor}
\end{equation}
where $\Gamma^M$'s and  $\omega_{abM}$'s are the Dirac matrices in eleven dimensions and the spin connection coefficients, respectively. The capital indices $M,N,...$ denote the eleven dimensional world coordinates and $a,b,...$ denote the tangent space coordinates in eleven dimensions. 

Using (\ref{killingspinor}) for the solution (\ref{ds11genM2}) with the transverse space (\ref{H1H2H}) or (\ref{BIX}) and the four-form field strength (\ref{4Field}), one finds 
\begin{eqnarray}
&&\left( 1-\Gamma ^{{t}{x}_{1}{x}_{2}}\right) \varepsilon =0, \label{proj012} \\
&&\left( 1-\Gamma ^{{t}{y}_{1}{y}_{2}}\right) \varepsilon =0, \label{proj034} 
\end{eqnarray}
that means at most one-fourth of the supersymmetry can be preserved due to the presence of two M2 branes.

Moreover from the first and second terms of (\ref{killingspinor}) for the solution (\ref{ds11genM2}) with the transverse space (\ref{H1H2H}), we get the following equations 
\begin{align}
\partial_{\eta}\varepsilon -\frac{1}{2} \Gamma^{\rho\eta}\varepsilon=0,&& \label{spEQ1} \\
\partial_{\psi}\varepsilon+\left[ \frac{n^2}{2(r+n)^2} \left( \Gamma^{{\psi}{r}} + \Gamma^{{\theta}{\phi}} \right) \right] \varepsilon =0,&&  \label{spEQ2} \\
\partial_{\theta}\varepsilon+\left[ \left( \frac{n}{4(r+n)}-\frac{1}{2} \right) \Gamma^{{r}{\theta}} -\frac{n}{4(r+n)}  \Gamma^{{\phi}{\psi}} \right] \varepsilon = 0,&& \label{spEQ3} \\
\partial_{\phi}\varepsilon+ \bigg[   \sin (\theta) \left(\frac{n }{4(n+r)} -\frac{1}{2}\right)\Gamma^{{r}{\phi}} + \cos (\theta) \left(\frac{n^2 }{4(n+r)^2} -\frac{1}{2}\right)\Gamma^{{\theta}{\phi}} + && \nonumber \\
+\frac{n  \sin (\theta) }{4(n+r)} \Gamma^{{\theta}{\psi}}  -\cos (\theta) \left(\frac{n^2 }{4(n+r)^2}\right)\Gamma^{{r}{\psi}}   \bigg] \varepsilon =0.&& \nonumber \\ \label{spEQ4}
\end{align}

We note that in equations (\ref{spEQ1})-(\ref{spEQ4}), only the subscript of derivative operator is a world coordinate. For notational simplification, we use the same letters for both world and tangent coordinates. The solution to equations \eqref{spEQ1}, \eqref{spEQ3} and \eqref{spEQ4} is 
\begin{equation}
\varepsilon=\exp \left[{\frac{\eta}{2}\Gamma^{{\rho}{\eta}}}\right] \exp \left[{\frac{\theta}{2}\Gamma^{{\phi}{\psi}}}\right] \exp \left[{\frac{\phi}{2}\Gamma^{{\theta}{\phi}}}\right]\tilde{\varepsilon},
\label{spSOL}
\end{equation}
where $\tilde{\varepsilon}$ is independent of $\eta, \theta$ and $\phi$. Substituting (\ref{spSOL}) in (\ref{spEQ2}) yields an equation for $\tilde{\varepsilon}$ that implies 
\begin{equation}
(1-\Gamma^{{\psi}{r}{\theta}{\phi}})\tilde{\varepsilon}=0,
\label{project3}
\end{equation}
if $\tilde{\varepsilon}$ is independent of coordinate $\psi$, too. 
The projection operator in (\ref{project3}) eliminates another half of the supersymmetry. So, only one-eighth of supersymetry is preserved for the system of two M2 branes, given by fields (\ref{ds11genM2}), (\ref{H1H2H}) and (\ref{4Field}). A similar calculation shows the system of two M2 branes with transverse space (\ref{BIX}) leads to three projection equations (\ref{proj012}), (\ref{proj034}) and (\ref{project3}). Hence the number of preserved sypersymmetries is the same as the system of two M2 branes with the transverse space (\ref{H1H2H}). Moreover, we notice the dimensional reduction to ten dimensions does not change the number of preserved supersymmetries. This is in agreement with the fact that the ten-dimensional metric (\ref{metri}) describes a system of three D branes and preserves four supersmmetries.

\section{Conclusions}
\label{sec:Summary}

By embedding self-dual curvature geometries in $D=11$ supergavity, we 
find new class of systems with two M2 branes.
to D=11 supergravity. 
The first set of solutions is exact and the analytical metric functions for two membranes are presented in equations
(\ref{MH2CSTAUB}), (\ref{H1TAUBF}) along with the four-form field strength (\ref{4Field}). The second set of solutions is approximate where the two membrane metric functions are given by (\ref
{H2TSOL}), (\ref{H1sol}). The four-form field strength again is given by (\ref{4Field}). Moreover the analytical continuation of separation constant in the first set of solutions gives a new set of solutions that the membrane metric function are given by (\ref{MH2ICS2}) and (\ref{HHtilde}).
For all of these solutions the brane functions are 
convolution of a decaying function with a damped
oscillating function. The weight function for all solutions can be found by comparing the convolution integral to the membrane metric function in appropriate limits in which the overall transverse space to membrane becomes flat. 
Dimensional reduction of these supergravity solutions to 10 dimensions provides realization of system of three D branes in type IIA string theory.
We explicitly solve the Killing spinor equation and show all the supergravity configurations preserve 1/8 of the supersymmetry in agreement with the system of three D branes after dimensional reduction.
It would be interesting to construct new solutions for the system of two M2 branes on other types of Bianchi space \cite{BianchiA} as the overall transverse space. \newline

\vspace{0.25cm}
{\bf Acknowledgments}\newline

This work was supported by the Natural Sciences and Engineering Research Council of Canada.\newline

\end{document}